\documentclass[%
 reprint,
 amsmath,amssymb,
 aps, pre,
]{revtex4-2}

\usepackage{graphicx}
\usepackage{dcolumn}
\usepackage{bm}
\usepackage{url}
\usepackage{overpic}
\usepackage{hyperref}

\begin{document}

\title{Wave patterns in two-dimensional networks of non-locally coupled oscillators with phase delay}

\author{Phillip Nimphius}
\author{Nariya Uchida}

\affiliation{%
Department of Physics, Tohoku University, Sendai 980-8578, Japan}%

\date{\today}%

\begin{abstract}
We studied the wave patterns in non-locally, repulsively coupled oscillators on a 2D lattice. The repulsive coupling is tuned by the phase delay $\alpha \pi$ and the wave patterns are found in the regime $\alpha \in \left[0.5, 1\right]$. We focused on the growth of orientationally correlated domains and found that the average total boundary size $\overline{|B|}$ obeys an approximate power law $\overline{|B|}\propto t^{-b}$ for $\alpha = 0.8$ and $\alpha =0.9$. In contrast, at $\alpha = 0.7$, the dynamics is disrupted by defect-mediated domain formation and domain splitting. The fitting-window dependence of the apparent exponent $b$, as well as the mean and standard deviation of the wave speed $c$, decreases with increasing $\alpha$, which is consistent with the linear stability analysis of the wave solution.
\end{abstract}

\maketitle

\section{\label{sec:Introduction}Introduction}

Large systems of coupled oscillators are ubiquitous in nature and technology, and they exhibit a wide variety of collective dynamical states, including synchronization, waves, defects, and spatiotemporal disorder \nolinebreak\cite{Kuramoto1984, Pikovsky2001, Strogatz2000}. A central problem in nonlinear dynamics is to understand how such systems self-organize from disordered initial conditions into coherent spatial and temporal patterns.

Nonlocally coupled phase oscillators provide a minimal framework for studying this problem. Kuramoto and Battogtokh showed that introducing a phase delay into nonlocal coupling can generate chimera states, in which coherent and incoherent regions coexist in a system of identical oscillators \nolinebreak\cite{Kuramoto2002}. 
Since then, nonlocally coupled oscillator networks have been studied
extensively, particularly in connection with chimera states and
related coherent--incoherent patterns~\cite{Abrams2004,Sethia2008,Omelchenko2011,Panaggio2015}.
Repulsively coupled oscillator networks have also been investigated in
one dimension, where they exhibit large-scale spatiotemporal patterns
distinct from those in the attractive regime
\nolinebreak\cite{Li2021}.

Two-dimensional networks have also been studied, mainly in the
attractive regime, where spiral-wave chimeras, spot and stripe
chimeras, twisted chimeras, multicore spiral chimeras, and related
coherent--incoherent structures appear
\nolinebreak\cite{ShimaKuramoto2004,Martens2010,Omelchenko2012,
PanaggioAbrams2013,Xie2015,Laing2017,
OmelchenkoWolfrumKnobloch2018}.
The stability of twisted states on two-dimensional oscillator lattices
has also been analyzed \cite{Goebel2021}, and topological defects in
spiral-wave chimera states have been characterized using
winding-number fields \cite{LiuUchida2026}.

In three-dimensional oscillator networks, chimera states can form
scroll waves, balls, tubes, and layers, as well as linked and knotted
filaments \cite{Maistrenko2015,LauDavidsen2016}.
Related spherical, layered, and cylindrical chimera patterns have also
been reported in networks of spiking-neuron oscillators
\cite{Kasimatis2018}.

More recently, partially coherent twisted states and their
long-wavelength stability were analyzed in a two-dimensional continuum
of nonlocally coupled phase oscillators with heterogeneous natural
frequencies in the attractive-coupling regime
\cite{omelchenko2025partially}.
In contrast, the repulsive regime in two dimensions has received much
less attention, although Kim et al. reported wave-like patterns in such
systems \nolinebreak\cite{Kim2004}.

The repulsive regime is particularly interesting because it does not simply lead to uniform phase synchronization. Instead, the phase delay destabilizes spatially uniform states and favors phase-gradient or twisted-wave configurations. Starting from random initial conditions, this can lead to the formation of multiple locally coherent traveling-wave domains with different orientations. The subsequent evolution of these domains raises several natural questions: how do these domains coarsen, what role do defects and domain boundaries play, and how does the phase delay parameter $\alpha$ control the stability of the resulting patterns?

In this paper, we study wave patterns in a two-dimensional lattice of nonlocally, repulsively coupled oscillators. We show that orientationally correlated traveling-wave domains emerge from random initial conditions and are separated by boundary regions containing arrays of topological defects. For sufficiently large phase delay, $\alpha \ge 0.8$, the total boundary size exhibits an approximate power-law decay, indicating coarsening of the wave domains. In contrast, for $\alpha = 0.7$, the coarsening dynamics is qualitatively different: coherent domains can split and new domains can emerge from defects, suggesting an instability-driven regime below a stability threshold. We also compare these numerical observations with a linear stability analysis of traveling-wave solutions and discuss anisotropic effects caused by the square lattice.

\section{\label{sec:model}Model}

We consider a square lattice of identical phase oscillators with nonlocal coupling.
The phase of the oscillator at site $(i,j)$ is denoted by $\theta_{ij}$, and its dynamics is given by
\begin{align}
\frac{d}{dt'}\theta_{ij}(t')=\omega-\frac{K}{N(R)}\sum_{0<|\Delta \bm{r}|\leq R}\sin\left[\theta_{ij}(t')-\theta_{mn}(t')+\alpha\pi\right],
\label{eq:model_original}
\end{align}
where $\Delta \bm{r}=(m-i,n-j)$ is the displacement vector from oscillator $(i,j)$ to oscillator $(m,n)$.
Here, $\omega$ is the natural frequency, $K>0$ is the coupling strength, $R$ is the coupling range, and $N(R)$ is the number of oscillators within the coupling range.
In this work, we use $R=6$, for which $N(R)=112$.

The phase delay is controlled by the parameter $\alpha$.
For $\alpha<1/2$, the coupling is effectively attractive near the spatially uniform state, whereas for $\alpha>1/2$ it is effectively repulsive.
We focus on the repulsive regime, where phase-gradient wave patterns are formed.

To remove the uniform rotation with frequency $\omega$, we introduce the phase variable in the rotating frame,
\begin{align}
\phi_{ij}(t') = \theta_{ij}(t') - \omega t' .
\label{eq:rotating_phase}
\end{align}
We also introduce the dimensionless time
\begin{align}
t = K t' .
\label{eq:dimensionless_time}
\end{align}
Then Eq.~\eqref{eq:model_original} becomes
\begin{align}
\frac{d}{dt}\phi_{ij}(t)=-\frac{1}{N(R)}
\sum_{0<|\Delta \bm{r}|\leq R}
\sin\left[
\phi_{ij}(t)-\phi_{mn}(t)+\alpha\pi
\right].
\label{eq:Kuramoto}
\end{align}
Eq.~\eqref{eq:Kuramoto} is the dimensionless equation used in the following analysis and simulations.


\section{\label{sec:lin}Linear stability analysis}

To understand the locally coherent wave domains observed in the simulations, we examine the linear stability of planar traveling-wave solutions of Eq.~\eqref{eq:Kuramoto}.
We consider an $L\times L$ periodic lattice and assume a $q$-twisted wave of the form
\begin{align}
\phi_{\bm r}^{(0)}(t)
=
\Omega t + \bm Q\cdot\bm r ,
\label{eq:twisted_wave}
\end{align}
where $\bm r=(i,j)$ denotes the lattice position,
\begin{align}
\bm Q = \frac{2\pi}{L}\bm q ,
\qquad
\bm q=(q_x,q_y)\in \mathbb{Z}^2 ,
\label{eq:wave_vector}
\end{align}
and $\Omega$ is the collective frequency.
Substituting Eq.~\eqref{eq:twisted_wave} into Eq.~\eqref{eq:Kuramoto}, we obtain
\begin{align}
\Omega(\bm Q) = -\frac{1}{N(R)} \sum_{0<|\Delta\bm r|\leq R} \sin\left[ -\bm Q\cdot\Delta\bm r+\alpha\pi \right].
\label{eq:Omega}
\end{align}
Since the interaction region is symmetric under $\Delta\bm r\to -\Delta\bm r$, this can also be written as
\begin{align}
\Omega(\bm Q)
=
-\frac{\sin(\alpha\pi)}{N(R)}
\sum_{0<|\Delta\bm r|\le R}
\cos\left(
\bm Q\cdot\Delta\bm r
\right).
\label{eq:Omega_symmetric}
\end{align}

We next perturb the traveling wave as
\begin{align}
\phi_{\bm r}(t)
=
\Omega t+\bm Q\cdot\bm r+u_{\bm r}(t),
\label{eq:perturbed_wave}
\end{align}
where the perturbation is expanded into Fourier modes,
\begin{align}
u_{\bm r}(t)
=
\sum_{\bm\kappa}
A_{\bm\kappa}
\exp\left(
i\bm\kappa\cdot\bm r
+
\Lambda_{\bm\kappa} t
\right),
\qquad
\bm\kappa=\frac{2\pi}{L}(n_x,n_y).
\label{eq:perturbation}
\end{align}
Linearizing Eq.~\eqref{eq:Kuramoto} with respect to $u_{\bm r}$ gives
\begin{align}
\dot u_{\bm r}
=
-\frac{1}{N(R)}
\sum_{0<|\Delta\bm r|\le R}
\cos\left[
-\bm Q\cdot\Delta\bm r+\alpha\pi
\right]
\left(
u_{\bm r}-u_{\bm r+\Delta\bm r}
\right).
\label{eq:linearized}
\end{align}
Therefore, the growth rate of the Fourier mode $\bm\kappa$ is
\begin{align}
\Lambda_{\bm\kappa}(\bm Q)
=
\frac{-1}{N(R)}
\sum_{0<|\Delta\bm r|\le R}
\cos\left(
\alpha\pi- \bm Q\cdot\Delta\bm r
\right)
\left(
1-e^{i\bm\kappa\cdot\Delta\bm r}
\right)
\label{eq:growth_rate}
\end{align}
The real part determines the linear stability of the traveling wave.
Using again the symmetry of the interaction region, we obtain
\begin{align}
\mathrm{Re}\,\Lambda_{\bm\kappa}(\bm Q)
=&
-\frac{\cos(\alpha\pi)}{N(R)} \nonumber \\
&\cdot \sum_{0<|\Delta\bm r|\le R}
\cos\left(
\bm Q\cdot\Delta\bm r
\right)
\left[
1-\cos\left(
\bm\kappa\cdot\Delta\bm r
\right)
\right].
\label{eq:real_growth_rate}
\end{align}
The zero mode $\bm\kappa=\bm 0$ corresponds to a uniform phase shift and is neutrally stable.
For all other perturbations, a traveling wave is linearly stable when
\begin{align}
\mathrm{Re}\,\Lambda_{\bm\kappa}(\bm Q)<0
\qquad
\text{for all } \bm\kappa\ne \bm 0 .
\end{align}

\begin{figure}
\centering
\includegraphics[width=0.45\textwidth]{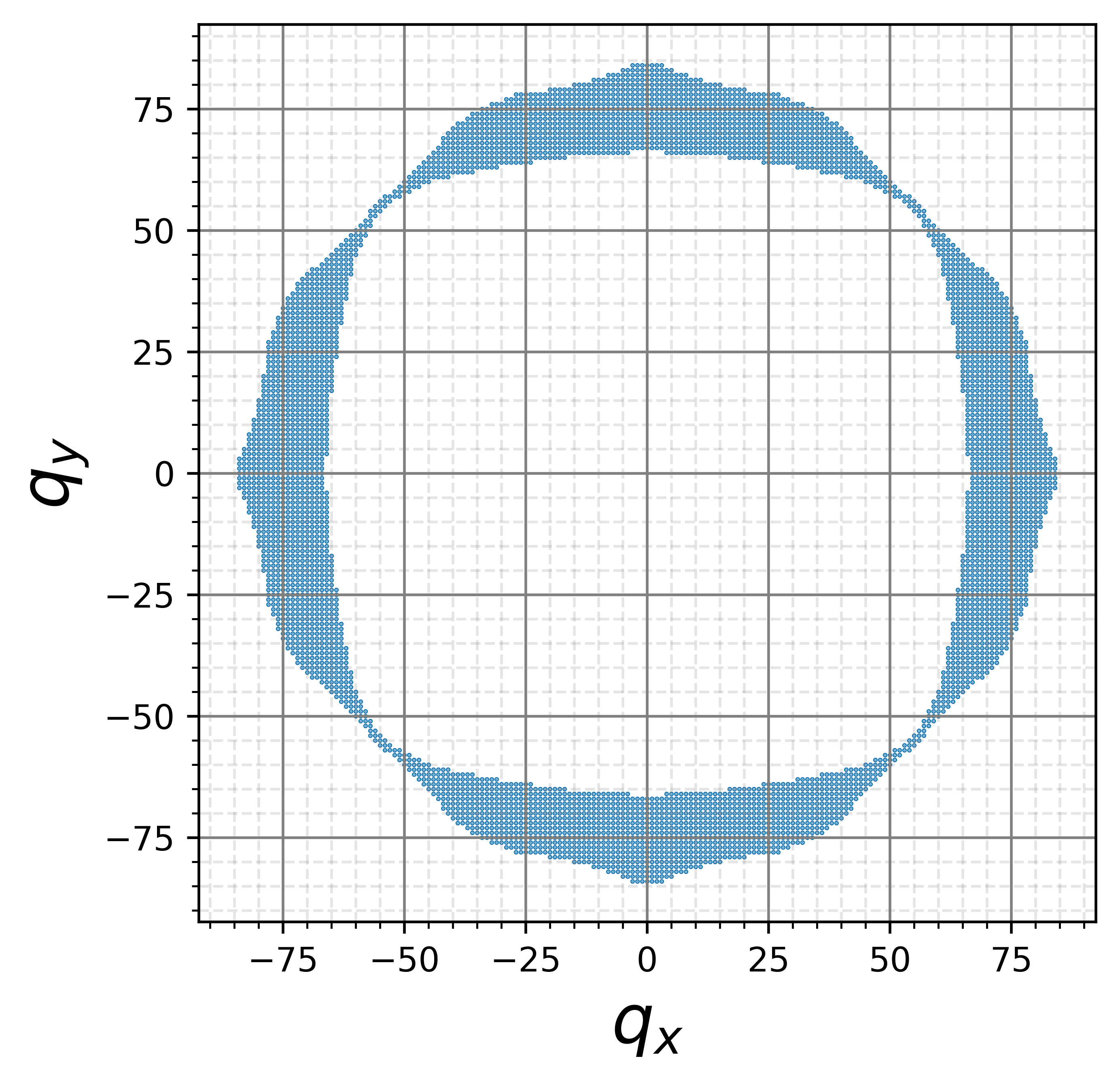}
\caption{\label{fig:stability} Linear stability diagram for q-twisted states for this system with $\alpha >0.5$. The blue dots mark vectors $(q_x, q_y) \in \left[-256, 255\right]$ for which every nonzero perturbation $\bm\kappa$ has a negative growth rate. The anisotropic shape reflects the discreteness and fourfold symmetry of the square lattice.}
\end{figure}

Eq.~\eqref{eq:real_growth_rate} shows that, within the repulsive regime $\alpha>1/2$, the set of stable wave vectors $\bm Q$ is independent of $\alpha$, because $\alpha$ appears only through the prefactor $\cos(\alpha\pi)$.
However, the magnitude of the growth or decay rate increases with $|\cos(\alpha\pi)|$.
Thus, stable traveling waves are more strongly damped against small perturbations as $\alpha$ approaches unity.
Fig. \ref{fig:stability} shows the stable $q$-twisted states obtained by direct numerical evaluation of Eq.~\eqref{eq:real_growth_rate}.
The anisotropic shape of the stable region reflects the discreteness and square symmetry of the lattice.

For comparison with the wave speeds measured in the simulations, we also define the speed of an ideal planar wave.
For the traveling wave in Eq.~\eqref{eq:twisted_wave}, the phase velocity in the direction normal to the wave fronts is
\begin{align}
c_{\rm th}(\bm Q)
=
\frac{|\Omega(\bm Q)|}{|\bm Q|}.
\label{eq:theoretical_wave_speed}
\end{align}
This theoretical speed is used only as a reference value, because the numerical patterns consist of finite domains with curved boundaries and topological defects, rather than perfectly uniform planar waves.

\section{\label{sec:level3}Numerical Results}
\subsection{Methods and Qualitative Description}
We integrated Eq.~\eqref{eq:Kuramoto} using the fourth-order Runge-Kutta method. We investigated $\alpha = 0.7$, $\alpha = 0.8$ and $\alpha = 0.9$. For each $\alpha$, 20 samples with randomized initial phases were simulated.

Each simulation was run for $2^{20}$ integration steps with time step $\Delta t=0.05$, corresponding to a final time $t_f = 52428.8$. The phase field was saved every 128 steps. As parameters we used the coupling range $R = 6$ and studied a network with $512 \times 512$ sites and periodic boundary conditions.

\begin{figure*}
\centering
\includegraphics[width=1\textwidth]{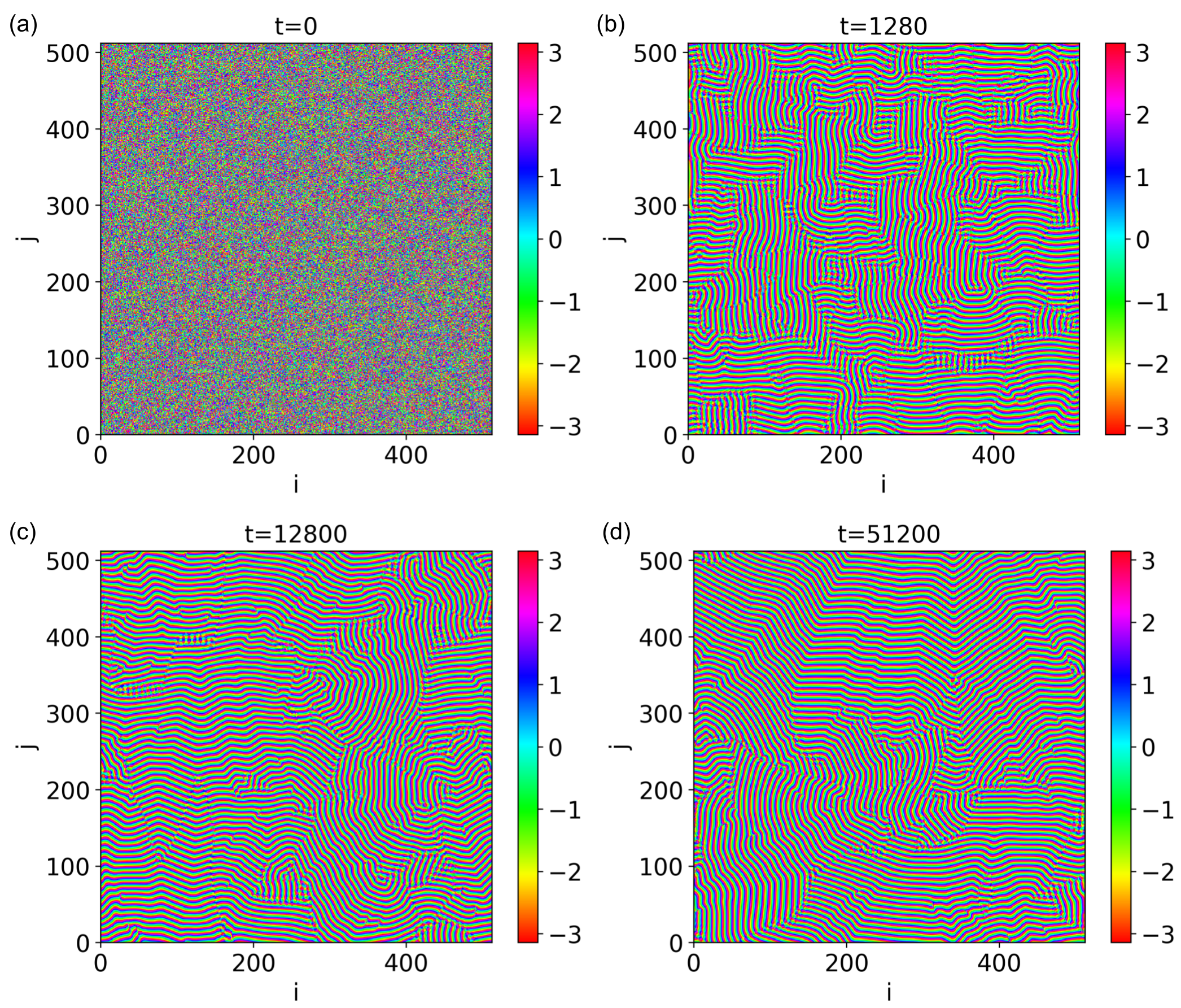}
\caption{\label{fig:phase_snapshots}
Snapshots of the oscillator phases $\phi_{ij}$ for a system with $\alpha=0.8$. Each pixel represents one oscillator on the two-dimensional lattice with the color representing the phase $\phi_{ij}$. The panels show (a) the random initial condition at $t = 0$, (b) $t=1280$,
(c) $t=12800$, and (d) $t=51200$.
Locally coherent traveling-wave domains emerge and coarsen over time.}
\end{figure*}

Fig. \nolinebreak\ref{fig:phase_snapshots} shows the time-evolution of a sample with $\alpha = 0.8$ and serves as an illustrative example as all studied samples show an early formation and expansion of locally coherent traveling-wave domains. Long-term behavior, however, depends more strongly on $\alpha$.
The corresponding phase evolutions for random initial conditions with $\alpha=0.7$, $0.8$, and $0.9$ are 
available as Supplementary Movies S1--S3 in the associated Zenodo repository~\cite{nimphius2026code}.

\subsection{Domain Boundaries}
\begin{figure*}
\centering
\includegraphics[width=1\textwidth]{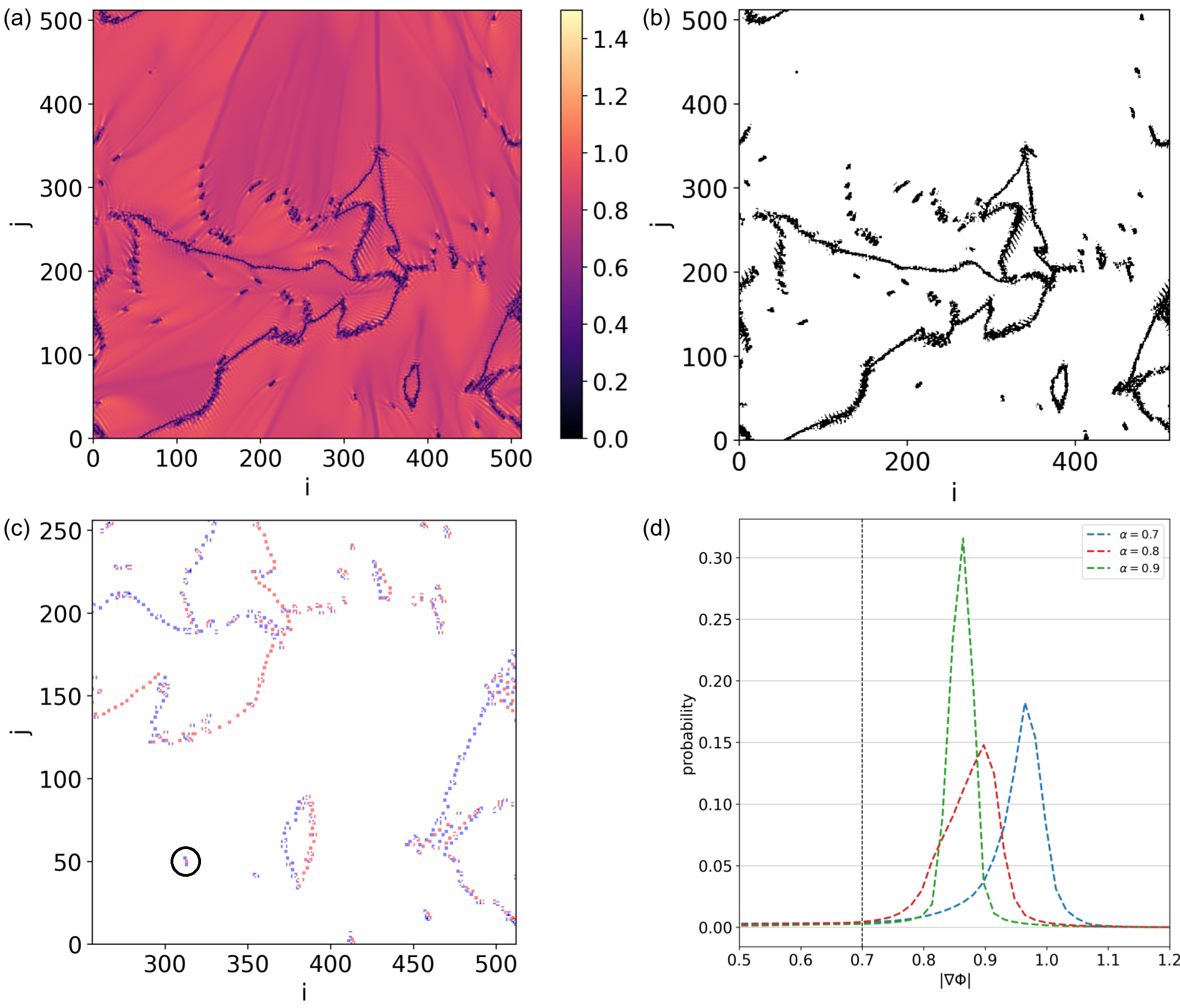}
\caption{\label{fig:boundary_defects}(a-c) show various snapshots of the same sample (with $\alpha = 0.8$) at the same time $t = 51 200$ as Fig. \ref{fig:phase_snapshots}(d). (a) shows the magnitude of the local average phase-gradient vector $|\langle {(\nabla \phi)}_{ij} \rangle|$. (b) shows the calculated boundary region of that snapshot in black. (c) shows the winding number $w_{ij}$ (blue positive, red negative). An isolated defect with total winding number $w = 1$ is marked with a black circle. (c) only displays values in the bottom right quadrant for better visibility. (d) shows the relative frequency of $|\langle {(\nabla \phi)}_{ij} \rangle|$ for $25 600 \le t \le 51 200$ for all $\alpha$. The threshold $\lambda = 0.7$ used to distinguish between boundary and non-boundary regions is marked by a dashed black line.}
\end{figure*}

To quantify the coarsening of these domains, we first need to distinguish between the domains $D$ and the boundary regions $B$. This is achieved by analyzing the local phase orientation. We define the discrete gradient vector
\begin{align}
    {(\nabla \phi)}_{ij} = \frac{1}{2} {(\phi_{i+1,j} - \phi_{i-1,j}, \phi_{i,j+1} - \phi_{i,j-1})}^T
\end{align}
at lattice site $(i,j)$. The phases of oscillators inside the boundary regions, however, are spatially uncorrelated, which implies that neighboring oscillators have phase gradients ${(\nabla \phi)}_{ij}$ with spatially uncorrelated magnitudes and angles. Since the domains are characterized by neighboring oscillators having similar phase gradients ${(\nabla \phi)}_{ij}$, we are able to distinguish the boundary regions $B$ by the magnitude of the local average of the gradient taken over the $3\times3$ neighborhood
\begin{align}
    \langle {(\nabla \phi)}_{ij} \rangle = \frac{1}{9}\sum_{|m-i|,|n-j|\leq 1} {(\nabla \phi)}_{mn}.
\end{align}
We use the histogram of $|\langle {(\nabla \phi)}_{ij} \rangle|$ seen in Fig.\ \nolinebreak\ref{fig:boundary_defects}(d) to determine an appropriate threshold $\lambda$, under which an oscillator is considered part of a boundary region $B = \{(i,j) \mid |\langle {(\nabla \phi)}_{ij} \rangle| < \lambda\}$. Based on the data, the threshold was set to $\lambda = 0.7$ for all $\alpha$. 
Figure~\ref{fig:boundary_defects} also shows that isolated defects are included in the boundary region, so that $|B|$ slightly overestimates the domain-interface area. The cluster analysis presented in Fig.~S2 of the Supplementary Material shows that the contribution of isolated defects remains minor over the analyzed time window, supporting the use of $|B|$ as a measure of the total boundary size.

\subsection{Winding Number}\label{sec:wind}
We observe two kinds of defects within the wave patterns. There are continuous arrays of defects at the boundary between domains and there are isolated defects. In order to quantify the defects we define the winding number $w_{ij}$
\begin{align}
    w_{ij} = \frac{1}{2\pi}\sum_{\gamma} \arg \left( e^{i(\phi_{\gamma+1} - \phi_\gamma)} \right)
\end{align}
where the index $\gamma$ enumerates the closed-loop path around the eight neighbor sites surrounding $(i,j)$. The total winding number $\sum_{ij} w_{ij}$ is constant.

Fig. \nolinebreak\ref{fig:boundary_defects}(c) shows the correspondence between the boundary region and the continuous array of defects. It also shows some isolated defects with a total winding number of $w_{\rm {defect}} \neq 0$. These defects persist until they merge with other defects. They can annihilate with an isolated defect with an opposite total winding number. Alternatively they can merge with the continuous array of defects between two domains. An interesting observation is that pairs of opposing winding numbers can spawn close to other defects with subsequent annihilation. The motion of isolated defects also suggests a possible effective short-range attraction that precedes annihilation, but a quantitative analysis is left for future research.

\subsection{\label{sec:subsec:Expansion}Domain Expansion}

\begin{figure}
\centering
\includegraphics[width=0.48\textwidth]{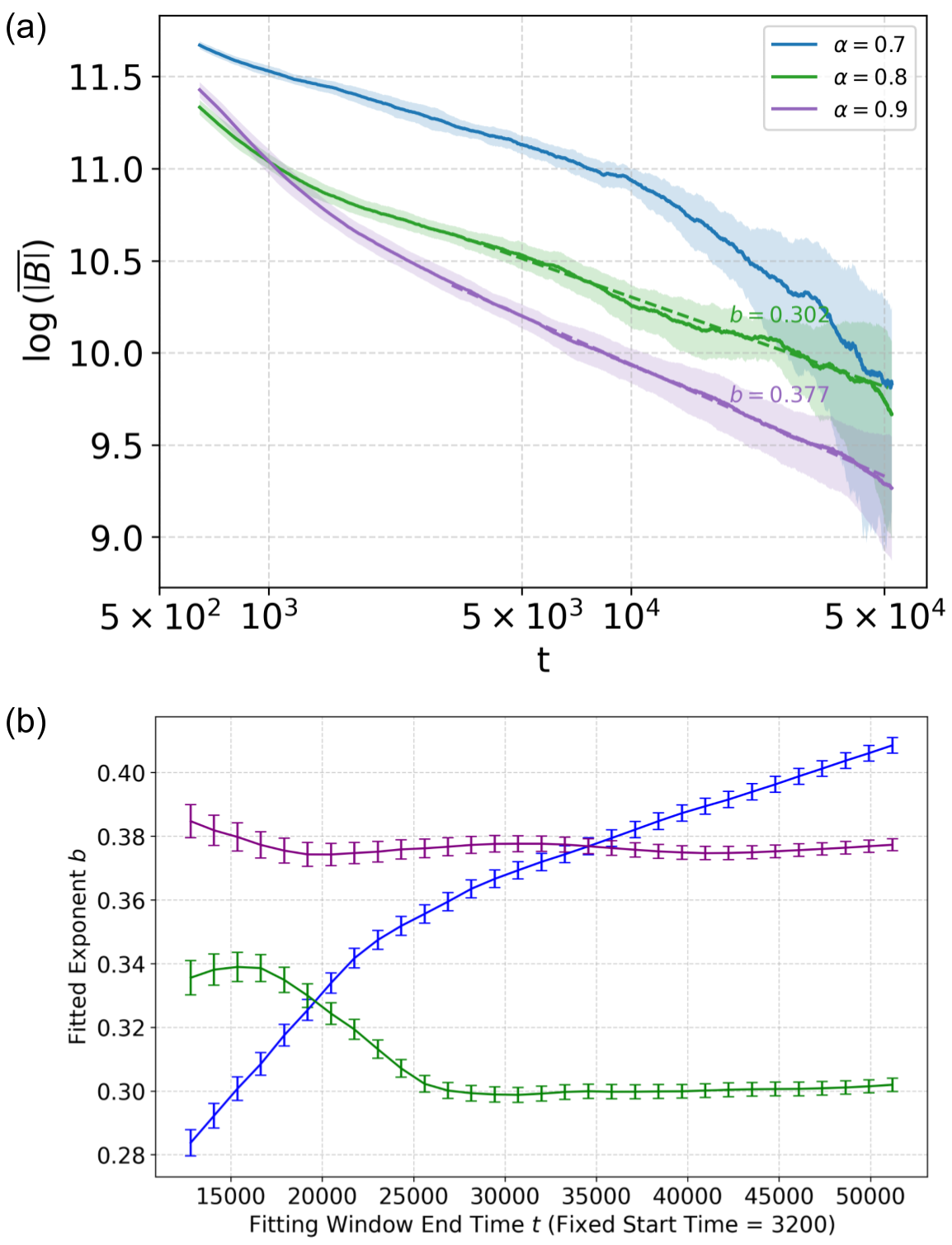}
\caption{\label{fig:boundary_decay}(a) shows the time evolution of the average total boundary size for each $\alpha$. The average is taken from all 20 samples for each $\alpha$. The shaded area indicates the sample-to-sample standard deviation. The range $3 200 \le t < 51 200$ has been fitted with $A t ^ {-b}$. The exponent is shown on the graph. The power-law decay is clearly visible for $\alpha = 0.8$ and $\alpha = 0.9$. The fitted exponents are $b_{0.8} = 0.30 \pm 0.04$ and $b_{0.9} = 0.377 \pm 0.008$. (b) displays the dependence of the fit on the end point of the fitting window (the start point is always $t = 3 200$). 
The fitted exponent for \(\alpha\ge0.8\) becomes relatively insensitive to the fitting-window end time for \(t_{\rm end}\gtrsim27500\), while it strongly depends on the fitting window for $\alpha = 0.7$. This supports power-law decay for $\alpha \geq 0.8$, but indicates different dynamics for $\alpha = 0.7$.
The error bars in panel (b) are the covariance-based statistical
uncertainties of the fitted exponents. The uncertainties quoted for
$b_{0.8}$ and $b_{0.9}$ combine this statistical uncertainty with the
standard deviation associated with the fitting-window dependence in
quadrature.
}
\end{figure}

To quantify the domain growth, we assume the characteristic domain size $a$ expands over time, following the power law $a(t) \propto t^{b}$. Thus we expect the total boundary length $l$ to follow a similar decrease with:
\begin{align}
    l(t) &= At ^ {-b},
\end{align}
where $A$ is a proportionality constant. We approximate the total boundary length $l$ using the number of lattice sites of the boundary region $|B|$. As shown in Fig.\ \nolinebreak\ref{fig:boundary_defects}(b), the boundary region $B$ exhibits quasi-one-dimensional qualities and has a width of several lattice spacings that does not change with $t$ or $\alpha$, making this a reasonable approximation. We calculate the ensemble average of the boundary size $\overline{|B|}(t)$ over 20 independent samples for each $\alpha$. The results are shown in Fig.\ \nolinebreak\ref{fig:boundary_decay}. We observe a power-law decay $\overline{|B|}(t) \propto t ^ {-b}$ for the datasets with $\alpha = 0.8$ and $\alpha = 0.9$. The case of $\alpha = 0.9$ shows fewer fluctuations compared to $\alpha = 0.8$. The exponent in the case of $\alpha = 0.7$ strongly depends on the fitting window, suggesting different dynamics compared to $\alpha \geq 0.8$.

\subsection{Wave Speed}
\begin{figure}
\centering
\includegraphics[width=0.48\textwidth]{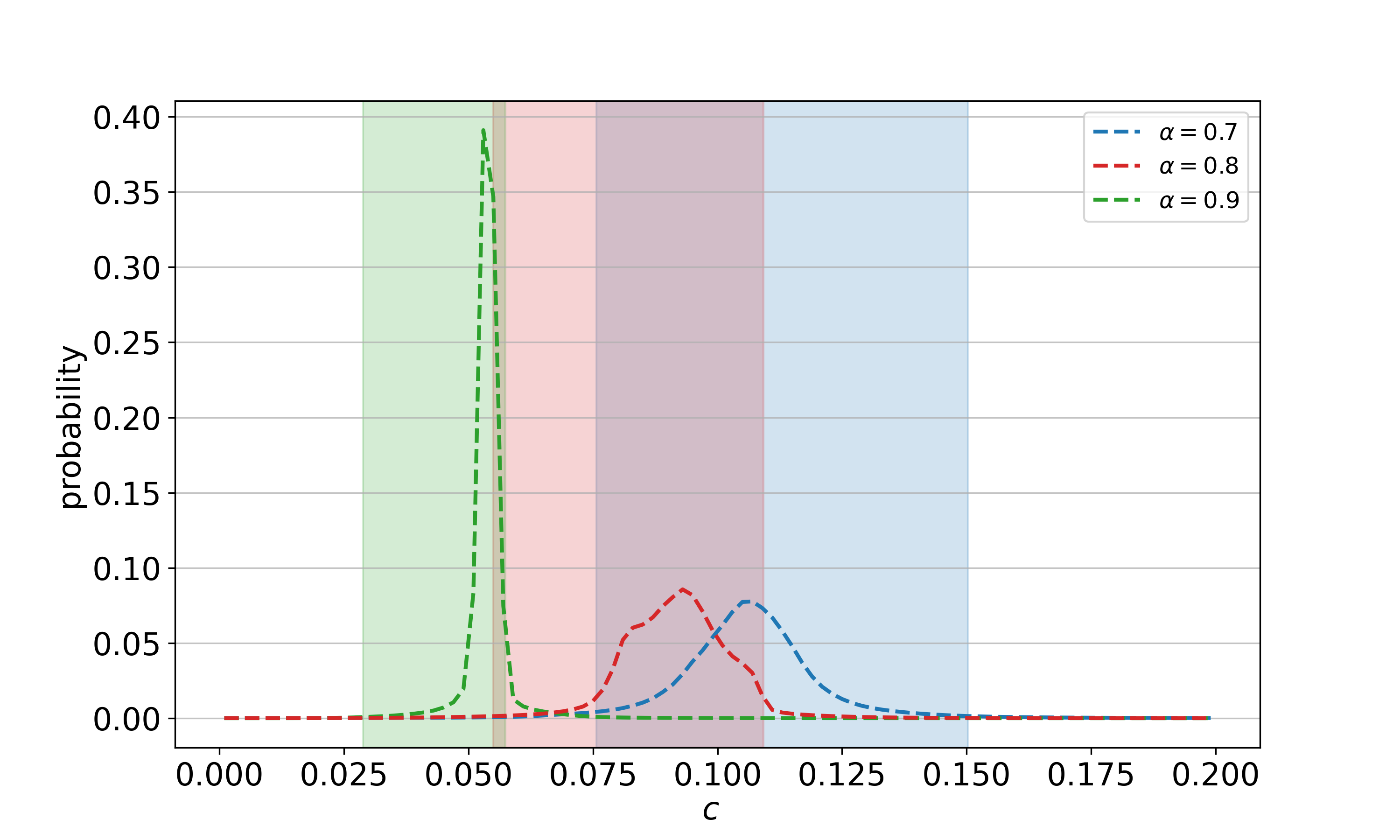}
\caption{\label{fig:wavespeed} Histogram for the absolute wave speed $c$ for the range $25 600 \le t \le 51 200$. Both mean and variance increase with decreasing $\alpha$. The wave speed is calculated from the gradient and the phase velocity $c_{ij} = \frac{|\Omega_{ij}|}{|\nabla\phi_{ij}|}$ with $\Omega_{ij} = \frac{d}{dt} \phi_{ij}(t)$. The shaded regions indicate theoretical wave-speed ranges for idealized planar waves obtained from the linear stability analysis, used here as reference ranges. The observed peaks fall within their respective ranges. 
Fig.~S1 in the Supplementary Material validates the numerical wave-speed calculation by comparing idealized planar-wave simulations with the theoretical planar-wave speed.
}
\end{figure}

We measure the wave speed at the node $(i,j)$:
\begin{align}
    c_{ij} &= \frac{|\Omega_{ij}|}{|\nabla\phi_{ij}|}\\
    \Omega_{ij} &= \frac{d}{dt} \phi_{ij}(t)
\end{align}
The results in Fig. \nolinebreak\ref{fig:wavespeed} show that wave speed depends on $\alpha$, with higher $\alpha$ having lower average wave speed $c$. We also observe higher fluctuations for lower $\alpha$ in the form of a larger variance of the wave speed $c$. The boundary region $B$ was excluded from the calculation of the wave speed and is neglected in the calculated histograms. 
From the linear stability analysis, we obtain the theoretical
collective frequency $\Omega(\bm Q)$ from Eq.~\eqref{eq:Omega}
and the phase-gradient vector $\bm Q$ for an ideal planar wave.
A comparison of these theoretical values with the observed values can be found in Fig. \ref{fig:wavespeed}.

\subsection{Circular Domain}\label{sec:Circle}

\begin{figure*}
\centering
\includegraphics[width=1\textwidth]{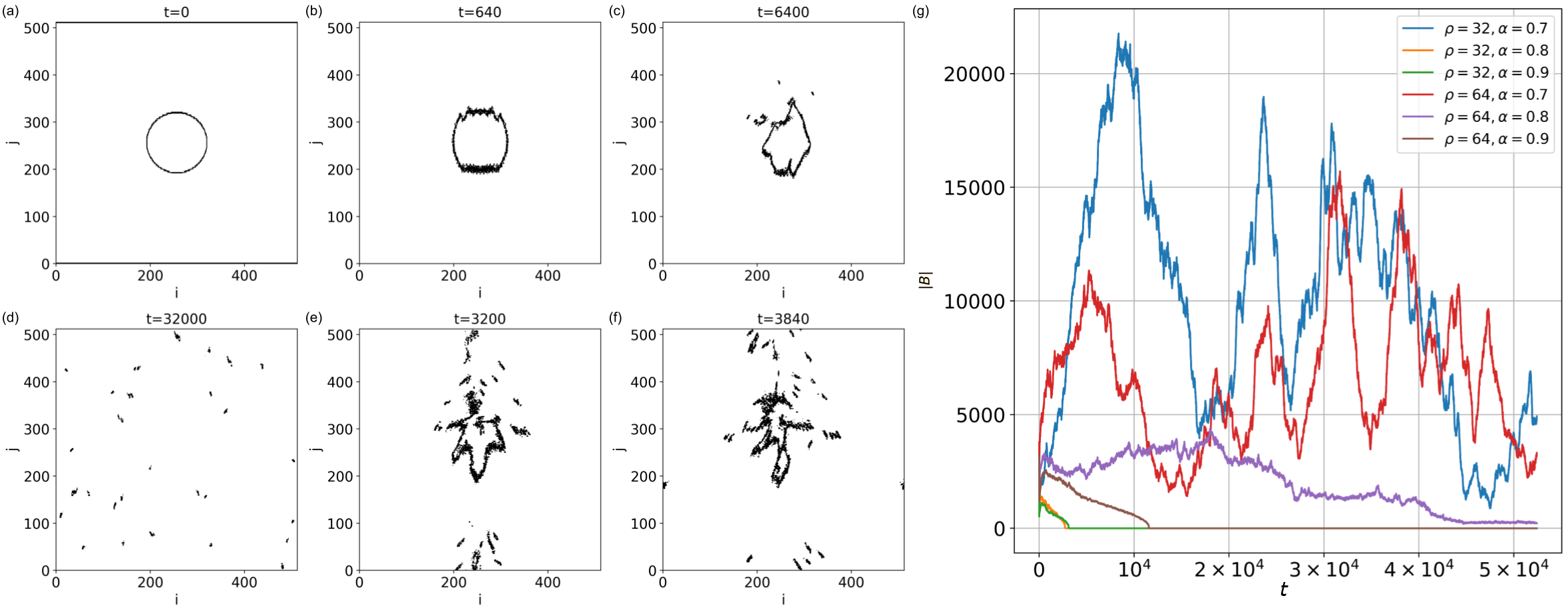}
\caption{\label{fig:Circle}(a-d) Time evolution of the boundary region of a circular domain with radius $\rho = 64$ and $\alpha = 0.8$ at (a) $t=0$, (b) $t=640$, (c) $t=6400$, and (d) $t = 32000$. The boundary deforms, shrinks, and eventually leaves only isolated defects. (e-f) A circular domain with the same radius, \(\rho=64\), but with $\alpha = 0.7$ at (e) $t = 3200$ and (f) $t = 3840$. In this case the boundary becomes unstable and splits into more complex structures. In the present simulations, such splitting is observed for $\alpha = 0.7$, but not for $\alpha = 0.8$ or $0.9$. (g) Time evolution of the boundary size $|B|(t)$ for initial radii $\rho=32,\ 64$. The cases $\alpha=0.8$ and $\alpha = 0.9$ evolve toward a single coherent wave domain, whereas $\alpha=0.7$ shows persistent fluctuations and does not stabilize during the simulated runtime.}
\end{figure*}

To study both the isotropy of the system and more controlled boundary configurations we simulated a circular domain embedded in a coherent domain with the opposite wave vector. Outside the circle, the phase was initialized as $\phi_{ij} = -\Delta i$, while inside the circle it was initialized as $\phi_{ij} = \Delta i$, with $\Delta = 0.85$. Thus the inner and outer domains have opposite wave vectors. The phase increment $\Delta = 0.85$ was chosen to match the typical phase gradient observed in the domains generated from random initial conditions. It is close to the periodic wave vector $\frac{2\pi \cdot 69}{512}$, and the small mismatch is treated as a small perturbation that does not affect the qualitative comparison in Fig.~\ref{fig:Circle}. 

We studied domains with radius $\rho = 32$ and $\rho = 64$ for each value of $\alpha$. Fig. \ref{fig:Circle}(a-d) shows snapshots of the time evolution. Fig. \ref{fig:Circle}(g) shows the time evolution of all simulated samples.
The corresponding phase evolutions for circular-domain initial conditions with $\alpha=0.7$, $0.8$, and $0.9$ are available as Supplementary Movies S4--S6 in the associated Zenodo repository~\cite{nimphius2026code}.

In all cases the circular shape starts to deform immediately, possibly influenced by both the wave vectors of the two domains and the geometry of the square lattice. During the initial deformation the domain retains symmetry. However, the symmetry breaks later, reflecting the instability of the circular boundary to small numerical or discretization perturbations.

After this development the long-term time evolution depends on $\alpha$. For $\alpha \geq 0.8$ the domain shrinks until it disappears entirely, sometimes leaving isolated defects behind.

For $\alpha = 0.7$ the domains never merge into a single system-spanning domain. Sometimes isolated defects serve as seeds for
the formation of new domains, which can leave new isolated defects when they collapse. This mechanism prevents the system from settling into a single coherent wave domain.

\section{\label{sec:level4}Discussion and Conclusion}

Previous studies of nonlocally coupled oscillators have mainly considered one-dimensional networks with attractive or repulsive coupling \cite{Kuramoto2002,Abrams2004,Li2021,LiUchida2022}, as well as two-dimensional systems in the attractive regime \cite{ShimaKuramoto2004,Martens2010}. 
More recently, Omel'chenko studied the existence and long-wavelength stability of individual partially coherent twisted states in a two-dimensional continuum with heterogeneous natural frequencies in the attractive-coupling regime \cite{omelchenko2025partially}. Kim et al. reported wave-like patterns in two-dimensional networks, including the repulsive-coupling regime, providing an early example of spatially organized states in this regime \cite{Kim2004}. The present study complements these works by focusing specifically on the repulsive-coupling regime and characterizing the spontaneous formation, coarsening, and defect-mediated splitting of multiple traveling-wave domains from random initial conditions in a finite lattice of identical oscillators.

The main result of this study is that the coarsening dynamics of wave domains changes qualitatively with the phase-delay parameter $\alpha$. For $\alpha=0.8$ and $0.9$, the boundary size $\overline{|B|}$ exhibits an approximate power-law decay over the simulated time range, consistent with coarsening of the wave domains. In contrast, for $\alpha=0.7$, the apparent scaling is disrupted by domain splitting and defect-mediated domain formation.

The expansion mechanism is distinct from other coarsening systems. Domains with characteristic size $R(t)$ in the Ising model and phase separation in fluids grow with $R(t) \propto t^b$ with $b = \frac{1}{2}$ and $b = \frac{1}{3}$  respectively \cite{bray1994theory}. They also have scalar order parameters. The characteristic distance between defects $R(t)$ in layer formation in nematic systems and Swift-Hohenberg systems grows with $R(t) \propto t^b$ with $b = \frac{1}{2}$ \cite{bray1994theory} and $b = \frac{1}{4}$ \cite{hou1997dynamical} ($b = \frac{1}{5}$ in the noiseless case), respectively. They have defect patterns similar to the isolated defects described in section \ref{sec:wind}, however they do not possess the continuous arrays of defects that are created by opposing wave patterns in the boundary region. All of these systems have very consistent exponents that are independent of boundary conditions, while in the present system the apparent exponent depends strongly on the phase delay parameter $\alpha$. Another difference to these systems is the anisotropic nature of the expansion as seen in the circular domain in Fig. \ref{fig:Circle} and the linear stability analysis in section \ref{sec:lin}.

Fig.~\ref{fig:Circle} further illustrates the reduced stability at lower $\alpha$. The results suggest a threshold between $\alpha=0.7$ and $\alpha=0.8$: for $\alpha=0.7$, new domains sometimes emerge from isolated defects and coherent domains can split into smaller domains, whereas such events are not observed for $\alpha=0.8$ or $\alpha=0.9$. This additional instability disrupts simple coarsening and the associated power-law scaling. Consistently, the average boundary size $\overline{|B|}(t)$ for $\alpha=0.7$ in Fig.~\ref{fig:boundary_decay} shows a marked change in its time dependence around $t=10^4$, after coherent wave domains have already formed. At this stage, domain splitting can create additional boundary regions, so the subsequent decay no longer follows a simple power law.

The observed change between $\alpha = 0.7$ and $\alpha = 0.8$ suggests a stability threshold or bifurcation-like change in the wave-domain dynamics. While our 2D repulsive system operates under different conditions, Maistrenko et al. \cite{Maistrenko2014} demonstrated that in non-locally coupled oscillators, shifting the phase delay parameter $\alpha$ can push the system through specific bifurcation routes, creating new chimera states. The observed changes in the behavior (emergence of new domains from isolated defects and the splitting of coherent domains) suggest the existence of a transition in the balance between the stabilizing and destabilizing forces within the system. Determining the nature of this transition requires a more detailed scan of $\alpha$ and is left for future work.

Future research could also focus on the effects of the lattice structure. The linear stability analysis shows anisotropic behavior for planar waves, likely influenced by the square lattice structure. A hexagonal or a random lattice structure may show very different behavior.

\section*{DATA AVAILABILITY}

The simulation and analysis code used in this study is openly available in Zenodo \cite{nimphius2026code}. The numerical data that support the findings of this study are available from the corresponding author upon reasonable request.

\bibliography{nimphius2026wave}

\end{document}